%% file: main.tex
\documentclass[journal]{IEEEtran}
\IEEEoverridecommandlockouts

\usepackage{cite}
\usepackage{amsmath,amssymb,amsfonts,amsthm}
\usepackage{graphicx}
\usepackage{textcomp}
\usepackage{xcolor}
\usepackage{cases}
\usepackage[utf8]{inputenc}
\def\BibTeX{{\rm B\kern-.05em{\sc i\kern-.025em b}\kern-.08em
    T\kern-.1667em\lower.7ex\hbox{E}\kern-.125emX}}
    
\usepackage[acronym,toc,shortcuts]{glossaries}
\makeglossaries
\input{glossaries.tex}

\usepackage{epstopdf}

\usepackage{hyperref}
\hypersetup{colorlinks=true,linkcolor=blue,}

\usepackage{dirtytalk}

\usepackage{enumitem}

\usepackage{algpseudocode}
\usepackage[linesnumbered,ruled,vlined]{algorithm2e}

\usepackage{ragged2e}

\usepackage{multirow}
\usepackage{dcolumn,booktabs}

\usepackage{subcaption}

\usepackage{empheq}

\begin{document}

\title{Energy-Efficient Coverage Enhancement of Indoor THz-MISO Systems: An FD-NOMA Approach}

\author{
    \IEEEauthorblockN{Omar~Maraqa\IEEEauthorrefmark{1}, Aditya S. Rajasekaran\IEEEauthorrefmark{2}\IEEEauthorrefmark{3}, Hamza U. Sokun\IEEEauthorrefmark{3}, Saad Al-Ahmadi\IEEEauthorrefmark{1}, Halim Yanikomeroglu\IEEEauthorrefmark{2}, \\Sadiq M. Sait\IEEEauthorrefmark{4}}\\
    \IEEEauthorblockA{\IEEEauthorrefmark{1}Department of Electrical Engineering and also the Center for Communication Systems and Sensing, King Fahd University of Petroleum \& Minerals, Dhahran-31261, Saudi Arabia}\\
    \IEEEauthorblockA{\IEEEauthorrefmark{2}Department of Systems and Computer Engineering, Carleton University, Ottawa, ON K1S 5B6, Canada}\\
    \IEEEauthorblockA{\IEEEauthorrefmark{3}Ericsson Canada Inc, Ottawa, ON K2K 2V6, Canada}\\
    \IEEEauthorblockA{\IEEEauthorrefmark{4}Center of Communications and IT Research, King Fahd University of Petroleum \& Minerals, Dhahran-31261, Saudi Arabia}
    
\thanks{$\copyright$ 2021 IEEE. Personal use of this material is permitted. Permission from IEEE must be obtained for all other uses, in any current or future media, including reprinting/republishing this material for advertising or promotional purposes, creating new collective works, for resale or redistribution to servers or lists, or reuse of any copyrighted component of this work in other works.}%

}

\markboth{Accepted in proceedings of 2021 IEEE International Symposium on Personal, Indoor and Mobile Radio Communications (PIMRC 2021)}%
{Maraqa \MakeLowercase{\textit{et al.}}: EE Coverage Enhancement of Indoor THz-MISO Systems: An FD-NOMA Approach}

\maketitle

\begin{abstract}
Terahertz (THz) communication is gaining more interest as one of the envisioned enablers of high-data-rate short-distance indoor applications in beyond 5G networks. Moreover, non-orthogonal multiple-access (NOMA)-enabled schemes are promising schemes to realize the target spectral efficiency, low latency, and user fairness requirements in future  networks. In this paper, an energy-efficient cooperative NOMA (CNOMA) scheme that guarantees the minimum required rate for the cell-edge users in an indoor THz-MISO communications network is proposed. The proposed cooperative scheme consists of three stages: (i) beamforming stage that allocates base-station (BS) beams to THz cooperating cell-center users using analog beamforming with the aid of the cosine similarity metric, (ii) user pairing stage that is tackled using the Hungarian algorithm, and (iii) power allocation stage for both the BS THz-NOMA transmit power and the cooperation power of the cooperating cell-center users, which are optimized sequentially. The obtained results quantify the energy efficiency (EE) of the proposed scheme and shed new light on the performance of multi-user THz-NOMA-enabled networks.  
\end{abstract}

\begin{IEEEkeywords}
Non-orthogonal multiple access (NOMA), user cooperation, full-duplex (FD), terahertz (THz) communication, energy efficiency (EE), beamforming, user pairing, power allocation. 
\end{IEEEkeywords}

\section{Introduction}

\IEEEPARstart{E}{}merging \ac{B5G} networks are witnessing revolutionary enhancement of data rate transmission for various innovative communication technologies and applications, including \ac{VR}, \ac{XR}, streaming $8$K videos over wireless links, etc. In addition to that, the immersion of \ac{umMTC} and \ac{IoE} in future wireless networks, that need ultra-reliable connectivity, are expected to become a reality in this decade~\cite{han2019terahertz}. Particularly, the vision of providing users with $1$ Tbps peak data rate, up to $10$ Gbps experience data rate, and almost perfect reliability is foreseen~\cite{8766143}. Nowadays, the trend of operating at high-frequency bands is apparent, for example, in \ac{5G} \ac{NR} standard (i.e., \ac{3GPP} release $16$) some \ac{mmWave} communications bands (i.e., between $24.25$ GHz to $52.6$ GHz) have been officially adopted. Nevertheless, for satisfying the aforementioned \ac{B5G} networks demands, these \ac{mmWave} communication bands might not be sufficient. Consequently,  operating at higher frequency bands, sub-\ac{THz} ($0.1$-$0.3$ THz) and \ac{THz} ($0.3$-$3$ THz) bands, where very large available contiguous bandwidth that spans from tens to even hundreds of GHz is  one of the promising solutions for this problem~\cite{han2019terahertz,8732419}. 

Traditionally, \ac{THz}-band applications have been restricted to sensing, imaging, and localization. This was due to the lack of compact high-power signal transmitters and high-sensitivity detectors~\cite{sarieddeen2019next} at this band. Recent advancements in \ac{THz} transceiver devices have paved the way for \ac{THz} communications~\cite{han2019terahertz}. Those advancements can be grouped into four technology paths, namely, electronic, photonic, integrated hybrid electronic-photonic, and plasmonic transceiver designs~\cite{sarieddeen2019next,han2019terahertz}.  Operating at sub-\ac{THz} and \ac{THz} frequency bands comes at the cost of high propagation losses. In~\cite{8732419}, Rappaport~\textit{et al.} highlight the additional challenges for wave propagation at frequencies beyond $100$ GHz, particularly that the free-space path loss and penetration loss increase significantly as we go to high frequencies, leading to shorter coverage areas. However, at these frequency bands, the wavelength and consequently the size of the antennas are very small, allowing for the use of highly directional antennas for ultra-precise beamforming~\cite{han2019terahertz}.



Over the past few years, researchers have started investigating \ac{NOMA} as a promising multiple access technique for accommodating more users in the same resource blocks~\cite{maraqa_nomasurvey2020,9303466}. Most of the research has focused on \ac{PD-NOMA} scheme which utilizes \ac{SC} at the transmitter and \ac{SIC} decoding at the receiver. In the \ac{PD-NOMA} scheme, multiple users share the same frequency and time resources and the channel gain differences are exploited at the receiver to separate the multiplexed user's signals using the \ac{SIC} decoding~\cite{maraqa_nomasurvey2020}, which offers significant improvements in connectivity and spectral efficiency. Also, the in-band user-assisted cooperative side links are proposed in the literature to enhance the spectral efficiency for the cell-edge users as well as the network energy efficiency~\cite{liu2015band}. 

In this paper, by utilizing the \ac{NOMA} principle along with \ac{FD} cooperative side links, we propose an energy-efficient user cooperative scheme that will guarantee the minimum required rate for cell-edge users in an indoor \ac{THz}-\ac{MISO} communications network. Particularly, due to the high propagation losses in \ac{THz} networks, cell-edge users might receive very weak \ac{THz} beams from the \ac{BS} and eventually suffer from a low data rate experience that is below their expected minimum required rate. Hence, with the aid of user-assisted cooperative side links and \ac{FD}-\ac{NOMA} scheme, this minimum required rate for cell-edge users can be guaranteed. The proposed cooperative scheme consists of three stages: (i) a beamforming stage that allocates beams to \ac{THz} cooperating cell-center users using analog beamforming with the aid of the cosine similarity metric, (ii) a user pairing stage of \ac{THz} cell-center users and user-assisted cell-edge users where a common matching algorithm, the Hungarian algorithm, that pairs users based on the Euclidean distance, is utilized and (iii) a power allocation stage where both the \ac{THz}-\ac{BS} \ac{NOMA} transmit power and the cooperation power of the cooperating cell-center users are optimized sequentially. Each of the aforementioned stages is designed to contribute to decreasing the consumed power in the network to make our proposed scheme energy-efficient.




The amount of work in the \ac{THz}-\ac{NOMA}-enabled networks is still limited~\cite{8824971,9115278,sabuj2020application}. In~\cite{8824971}, in order to maximize the network throughput, the authors solved an optimization problem that involves beamforming, power allocation, and sub-band assignment. Through numerical simulations, the authors demonstrated the superiority of the proposed scheme compared to the THz-OMA counterpart. In~\cite{9115278}, a downlink heterogeneous \ac{THz}-\ac{NOMA} cache-enabled system with imperfect \ac{SIC} is proposed. An \ac{EE} maximization problem that includes user clustering, beamforming, and power allocation is formulated and then solved by dividing the original problem into three sub-problems. Through simulations, the authors show that the proposed enhanced K-means algorithm for user clustering and the proposed alternating direction method of multipliers for power allocation can achieve higher \ac{EE} performance for the proposed network. In~\cite{sabuj2020application}, a machine-type communication downlink sub-\ac{THz}-\ac{NOMA} enabled system to enable massive connectivity of devices has been analyzed. An \ac{EE} maximization problem has been solved via unconstrained then constrained optimization. Through simulations, the authors show that the proposed system attains better performance compared to its counterpart in the mmWave band. There is no existing work that combines multi-user indoor \ac{MISO}-\ac{NOMA} in \ac{THz} band with user-assisted cooperative side links for coverage enhancement of cell-edge users.


The main contributions can thus be summarized as follows;
\begin{itemize}
    \item Proposing an energy-efficient user-assisted full-duplex cooperative \ac{NOMA} scheme that guarantees the minimum required rate for cell-edge users in multi-user indoor \ac{THz}-\ac{MISO} systems.
    \item Developing the proposed scheme in three stages: (i) analog beamforming stage with the aid of the cosine similarity metric, (ii) user pairing stage for \ac{THz} cooperating cell-center users and cell-edge users, and (iii) a power allocation stage for both the \ac{BS} transmit power and the cooperation power of the cooperating cell-center users.
    \item Investigating the effect of changing different network parameters through simulations, such as, the \ac{BS} transmit power levels and cell-edge users' minimum required rate on the \ac{EE} of the network. The obtained results are bench-marked with the \ac{mmWave}-\ac{NOMA} counterpart scheme.
\end{itemize}

The remainder of this paper is organized as follows. In Section~\ref{Section:System Model and Problem Formulation}, we provide the channel model and the system model of the proposed \ac{THz}-\ac{MISO} \ac{NOMA}-enabled scheme. The details of the proposed scheme are presented in Section~\ref{Sec: The Proposed Energy-Efficient CNOMA Scheme}. Simulation results are provided in Section~\ref{Section:Results and discussions}. Finally, paper conclusions and future research directions are given in Section~\ref{Section:Conclusions}.
 
\begin{figure}[t!]
\centering
\includegraphics[width=0.35\textwidth]{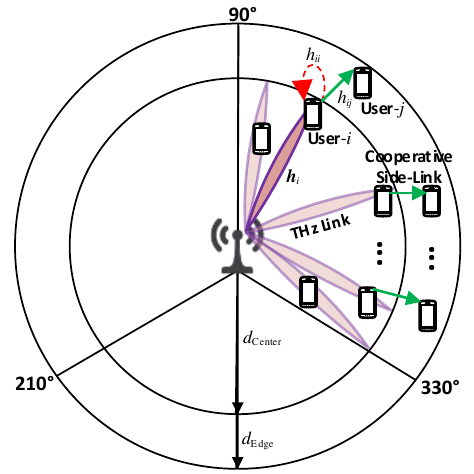}
\caption{An illustration of the proposed multi-user indoor \ac{THz}-\ac{MISO} \ac{NOMA}-enabled scheme with user-assisted cooperative side links.}
\label{fig: THz NOMA system model}
\vspace{-1.5em}
\end{figure}

\section{Channel Model and System Model}
\label{Section:System Model and Problem Formulation}

The system model of the proposed multi-user indoor \ac{THz}-\ac{MISO} \ac{CNOMA} scheme is depicted in Fig.~\ref{fig: THz NOMA system model}. Consider a downlink system with a single \ac{BS} equipped with one \ac{RF} chain and $N$-antenna phased arrays serving (i) $K_{\textnormal{Center}}$ cell-center users through \ac{ABF} and (ii) $K_{\textnormal{Edge}}$ cell-edge users through user-assisted full-duplex \ac{DF} cooperative side links by a subset of $K_{\textnormal{Coop}} \subset K_{\textnormal{Center}}$ cooperating users. All the users are equipped with a single antenna element. To facilitate the investigation of the proposed \ac{CNOMA} scheme, this paper assumes that $K_{\textnormal{Coop}}$ equals $K_{\textnormal{Edge}}$. In this work, due to the requirement of low power consumption and hardware cost, we consider the \ac{ABF} structure in the \ac{BS}, which utilizes the \ac{SPS} implementation~\cite{xiao2019user}. Also, for decreasing the computational complexity, it is assumed that the \ac{BS} has full \ac{CSI} and distance information for all users in the network, this assumption is also adopted by~\cite{8824971,9115278}. In the \ac{BS}, each antenna is driven by a \ac{PS} and a \ac{PA}. The elements of the precoding vectors are complex numbers, its phase is controlled by the phase shifters and its modulus is controlled by the power amplifiers~\cite{xiao2019user}.

\subsection{Indoor THz Channel Model}

In this subsection, we discussed the indoor \ac{THz} channel model. In the \ac{THz} band, the high path loss resulting from spreading loss and molecular absorption loss limits the scattering phenomenon. This makes \ac{THz} communication links sensitive to obstacles blockage such as walls, and leads to large path loss difference between the direct paths and blocked paths. For example, in an indoor \ac{THz} system, Priebe and Kurner~\cite{6574880} reported that when the \ac{LoS} exists, the first-order reflections are attenuated on average more than $10$ dB, and the second-order reflections are attenuated on average more than $20$ dB. From this, one can conclude that the \ac{NLoS} paths of \ac{THz} channels are very limited. Therefore, in this work similar to~\cite{9115278}, we consider the \ac{LoS} paths only. Hence, the channel gain vector of the $i$-th user on the $k$-th user pair can be expressed as
\begin{equation} \label{eq: THz channel}
\mathbf{h}_{k,i}=  \sqrt{N} ( \sqrt{\frac{1}{\mathcal{PL}(f,d)}}  \Omega  \boldsymbol{a} ( \theta_{k,i} )),
\end{equation}
\begin{equation} \label{eq: Path loss}
\begin{split}
\mathcal{PL}(f,d) &= \mathcal{L}_{\textnormal{spread}}(f,d) \mathcal{L}_{\textnormal{abs}}(f,d)=  \big( \frac{4 \pi f d}{c} \big)^2 e^{k_{\textnormal{abs}}(f)d},
\end{split}
\end{equation}
\noindent where $\mathcal{PL}(f,d)$ denotes the path loss incurred by the \ac{THz} signal that is transmitted from the \ac{BS} on frequency $f$ to the $i$-th user that is $d$ meters apart from the \ac{BS}. $\Omega$ and $\boldsymbol{a} ( \theta_{k,i})$ denote the antenna gains of the \ac{BS} and the array steering vector toward user-$i$ of the $k$-th user pair, respectively. Equation \eqref{eq: Path loss} shows the two multiplicative factors that contribute to the path loss of \ac{THz} signals, namely, the spreading loss $\mathcal{L}_{\textnormal{spread}}$ and the molecular absorption loss $\mathcal{L}_{\textnormal{abs}}$. The term $k_{\textnormal{abs}}(f)$ denotes the frequency-dependent absorption coefficient for different isotopologues of water vapor molecules. It is worthy to note that, in a regular medium, the main contributor of the total absorption loss comes from water vapor molecules and its loss far exceeds the contribution of other air molecules~\cite{9115278}. The speed of light is denoted by $c$. The term $\boldsymbol{a} (\theta_{k,i})$ represents the array steering vector for \ac{ULA} and can be represented as~\cite{9115278}
\begin{equation} \label{eq: steering vector}
\boldsymbol{a} (\theta_{k,i}) =\frac{1}{\sqrt{N}} [1, ..,  e^{j \pi [n \sin(\theta_{k,i})]}, .., e^{j \pi [(N-1) \sin(\theta_{k,i})]} ]^{T}, 
\end{equation}
\noindent where $\theta_{k,i}$ denotes the physical angle-of-departure of the \ac{THz} beam. The cooperative side-link channel model also follows the one in \eqref{eq: THz channel}; since in-band user-assisted cooperative communications paradigm is adopted~\cite{liu2015band,9201439}. 

\subsection{System Model and Formulation}
\label{Section: System Model and Formulation}

In this subsection, we present the details of the proposed multi-user indoor \ac{THz}-\ac{MISO} \ac{NOMA}-enabled system and its energy-efficiency formulation. For mathematical convenience, we focus the analysis on the signaling within one \ac{NOMA} user pair (i.e, the cooperating cell-center user-$i$ and cell-edge user-$j$), the remaining user pairs can be analyzed similarly, since in this work, each \ac{THz} cooperating cell-center user is served by one precoding vector on an orthogonal channel which can be a time slice or a sub-carrier frequency channel. With the adoption of the \ac{PD-NOMA} scheme, the \ac{BS} transmits superimposed signal towards the $k$-th user pair is
\begin{equation} \label{eq: Tx to user-i in THz band}
 \mathbf{s}_{k}=\mathbf{w}_k s_k=\mathbf{w}_k (\sqrt{\beta_{k,i}p_k} s_{k,i} + \sqrt{\beta_{k,j}p_k} s_{k,j}),
\end{equation}
\noindent where $\mathbf{w}_k$ represents the precoding vector formed toward $k$-th user pair, $s_{k}$ represents the \ac{SC} signal for the $k$-th user pair, $\beta_{k,i}$ and $\beta_{k,j}$ denote the power fractions assigned by the \ac{BS} for cooperating cell-center user-$i$ and cell-edge user-$j$, respectively. $s_{k,i}$ and $s_{k,j}$ denote the transmit messages of cooperating cell-center user-$i$ and cell-edge user-$j$. $p_k$ denotes the \ac{BS} transmit power for the $k$-th user pair (i.e., per-channel use).

As depicted in Fig.~\ref{fig: THz NOMA system model}, the received signal at the $i$-th cooperating cell-center user on the $k$-th user pair while considering \ac{FD} relaying mode for the cooperative side link is
\begin{equation} \label{eq: Rx to user-i in THz band}
\begin{split}
y_{k,i}^{\textnormal{Coop}} & = \underbrace{\mathbf{h}_{k,i}^{H} \mathbf{w}_{k} \sqrt{\beta_{k,i}p_k} s_{k,i}}_{\textnormal{Desired Signal for User}_i} + \underbrace{\mathbf{h}_{k,i}^{H} \mathbf{w}_{k}  \sqrt{\beta_{k,j} p_k} s_{k,j}}_{\textnormal{Intended Signal for User}_j}\\ 
& + \underbrace{h_{k,ii} \sqrt{p_{k,u}} \hat{s}_{k,ii} }_{\textnormal{Self Interference}} + \underbrace{n_{k,i}}_{\textnormal{Noise}},
\end{split}
\end{equation}
\noindent where $h_{k,ii}$ represents the \ac{SI} channel gain for the $i$-th user of the $k$-th user pair. $\hat{s}_{k,ii}$ represents the \ac{SI} signal of the full-duplex link that arrives at the $i$-th user. $p_{k,u}$ represents the cooperative transmit power of the $i$-th user of the $k$-th user pair. $n_{k,i}$ denotes the noise seen at user-$i$ of the $k$-th user pair. The second term in~\eqref{eq: Rx to user-i in THz band}, represents the intended signal for the $j$-th user of the $k$-th user pair that will be extracted by performing the \ac{SIC} decoding procedures. On the other hand, the received signal at the $j$-th cell-edge user on the $k$-th user pair can be expressed as  
\begin{equation} \label{eq: Rx to user-j in cooperative}
 y_{k,j}^{\textnormal{Edge}}= \underbrace{h_{k,ij}\sqrt{p_{k,u}} \hat{s}_{k,ij}}_{\textnormal{Desired Signal}}+ \underbrace{n_{k,j}}_{\textnormal{Noise}},
\end{equation}
\noindent where $h_{k,ij}$ denotes the cooperative side-link channel gain between user-$i$ and user-$j$ of the $k$-th user pair. $\hat{s}_{k,ij}$ represents the transmit signal from user-$i$ to user-$j$ of the $k$-th user pair. $n_{k,j}$ denotes the noise seen at user-$j$ of the $k$-th user pair. Now, we can express the \ac{SINR} when decoding the $j$-th user signal at the $i$-th user of the $k$-th user pair as
\begin{equation} \label{eq: SINR of the cooperative user at the THz user}
  \gamma_{k,j}^{i,\textnormal{Coop}}=  \frac{\beta_{k,j}p_k |\mathbf{h}_{k,i}^{H} \mathbf{w}_{k} |^2}{\beta_{k,i}p_k |\mathbf{h}_{k,i}^{H} \mathbf{w}_{k}|^2 + p_{k,u} \kappa  |h_{k,ii}|^2 +  \sigma_{k,i}^2},
\end{equation}
\noindent where $\kappa$ denotes the \ac{SI} parameter, and its value ranges between $[0,1]$. When $\kappa=0$, this refers to the perfect \ac{SI} cancellation scheme. On the other hand, when $\kappa=1$, this refers to the scheme where there is no \ac{SI} cancellation at the $i$-th cooperating cell-center user. $\sigma_{k,i}^2$ denotes the noise power. After performing the \ac{SIC} decoding procedures, the \ac{SINR} when user-$i$ decodes its own signal can be expressed as
\begin{equation} \label{eq: SINR of the THz user at the THz user}
  \gamma_{k,i}^{i, \textnormal{Coop}}=  \frac{\beta_{k,i}p_k |\mathbf{h}_{k,i}^{H} \mathbf{w}_{k}|^2}{p_{k,u} \kappa  |h_{k,ii}|^2 + \sigma_{k,i}^2}.
\end{equation}
\indent The \ac{SNR} seen at the cell-edge user-$j$ on the $k$-th user pair can be represented as
\begin{equation} \label{eq: SINR of the cooperative user}
  \gamma_{k,j}^{j,\textnormal{Edge}}=  \frac{p_{k,u} |h_{k,ij}|^2}{\sigma_{k,j}^2}.
\end{equation}
\indent At this point, we can represent the achievable rate of decoding user-$j$ signal at user-$i$ and the achievable rate when user-$i$ decodes its own signal as
\begin{equation} \label{eq: Rate of the weak at strong user}
  R_{k,j}^{i,\textnormal{Coop}}=W \log_2 (1 + \gamma_{k,j}^{i,\textnormal{Coop}}),
\end{equation}
\begin{equation} \label{eq: Rate of the THz user}
  R_{k,i}^{\textnormal{Coop}}=W \log_2 (1 + \gamma_{k,i}^{i,\textnormal{Coop}}),
\end{equation}
\noindent where $W$ is the available contiguous bandwidth of the \ac{THz} channel window. According to~\cite{4205048}, the achievable rate when user-$j$ decodes its own signal can be expressed as 
\begin{equation} \label{eq: Rate of the cooperative user}
  R_{k,j}^{\textnormal{Edge}}= \textnormal{min} \{W \log_2 (1 + \gamma_{k,j}^{i,\textnormal{Coop}}), W \log_2 (1 + \gamma_{k,j}^{j,\textnormal{Edge}}) \}.
\end{equation}
\indent In the following, we express the energy efficiency of the system that can be defined as \say{the ratio between the achievable sum rate and the total power required to achieve this rate (Bits/Joule)}~\cite{9132716} for the $K_\textnormal{Coop}$ orthogonal channels
\begin{equation} \label{eq: GEE formula}
\begin{split}
  \textnormal{EE} &= \frac{ \sum_{k=1}^{K_\textnormal{Coop}} (R_{k,i}^{\textnormal{Coop}} +  R_{k,j}^{\textnormal{Edge}})/K_\textnormal{Coop}}{\xi (( \sum_{k=1}^{K_\textnormal{Coop}} p_k + p_{k,u})/K_\textnormal{Coop}) + P_{\textnormal{loss}}}\\
  &= \frac{ \sum_{k=1}^{K_\textnormal{Coop}} (R_{k,i}^{\textnormal{Coop}} +  R_{k,j}^{\textnormal{Edge}})}{\xi ( \sum_{k=1}^{K_\textnormal{Coop}} p_k + p_{k,u}) + K_\textnormal{Coop} P_{\textnormal{loss}}},
\end{split}
\end{equation}
\noindent where $\xi$ is the power amplifiers' inefficiency at the \ac{THz} \ac{BS} as well as at the cooperating cell-center users. $P_{\textnormal{loss}}$ is the circuit power consumption and can be expressed as~\cite{9115278}
\begin{equation} \label{eq: P_loss}
 P_{\textnormal{loss}}=P_B + N_{RF} P_{RF} + N_T P_A + N_T P_P,
\end{equation}
\noindent where $P_B$ is the base-band power consumption, $N_{RF}$ is the number of \ac{RF} chains at the \ac{BS}, $P_{RF}$ is the \ac{RF} chain power consumption, $N_T$ is the number of power amplifier/phase shifters at the \ac{BS}, $P_A$ is the power consumption of each power amplifier, and $P_P$ is the power consumption of each phase shifter. 
 
\section{The Proposed Energy-Efficient CNOMA Scheme}
\label{Sec: The Proposed Energy-Efficient CNOMA Scheme}

In this section, we discuss in detail the proposed energy-efficient cooperative communication scheme. A summary of the proposed scheme is provided in Algorithm~\ref{Tab: A summary of the proposed cooperative scheme}.


\subsection{Cooperating Cell-center User Scheduling Stage: Analog Beamforming with the Cosine Similarity Metric}
\label{subsection: Cell-center User Scheduling: Analog Beamforming with the Cosine Similarity Metric}

Typically, in small-cell and indoor-cell deployments, a low hardware cost and power consumption are essential~\cite{9264161}. Hence, \ac{ABF} is adopted in this work to meet such demand. The \ac{BS} applies \ac{ABF} that has a fixed set of beams uniformly distributed over the sector coverage area. In this work, the indoor cell is assumed to have three sectors of $120^{\circ}$ each. With one \ac{RF} chain available at the \ac{BS}, only one beam can be transmitted at a time, which we equate to forming one beam to serve one \ac{NOMA} user pair per channel use. Since we use \ac{ABF} that can only generate one beam at a time, we use a time-division strategy to alternate between the different user pairs. Fig.~\ref{fig: THz NOMA system model} illustrates the entire angular coverage region that is divided into three sectors. In here, we consider a specific sector, $\bar\theta$, from $-\pi/6$ to $\pi/2$ or $330^{\circ}$ to $90^{\circ}$, other sectors can be analyzed similarly. In this sector, the $120^{\circ}$ area is covered by a set of $B+1$ beams. Each beam-$b$ of the available beams has the following steering vector~\cite{9264161}:
\begin{equation}\label{eq:Beams}
\mathbf{w}_b = \boldsymbol{a}(\bar\theta_b), \forall b \in [0,B],
\end{equation}
\noindent where the parameter $\bar\theta_b$ is
\begin{equation}\label{eq:Beam-b}
\bar\theta_b = -\pi/6 + (b \times \frac{2 \pi}{3 B}).
\end{equation}
\indent In this way, this entire sector region is divided into $B$ equal angles, effectively forming a set of $B+1$ beams. The $B+1$ beams can be thought of as a choice of $B+1$ different steering vectors based on~\eqref{eq:Beams}, such that collectively, the steering vectors of the $B+1$ candidate precoding vectors uniformly cover the entire sector region of $\bar\theta = -\pi/6$ to $\pi/2$. 

The cosine similarity metric is utilized to determine the level of correlation among \ac{THz} cooperating cell-center users and the \ac{BS} available beams that are formed through analog beamforming. Several works in mmWave-NOMA systems have used the cosine similarity metric to determine the correlations among the users channels~\cite{8454272} or between the users channels and the fixed beams~\cite{9264161} and this concept can be similarly used here in the context of \ac{THz}-\ac{NOMA} systems. In particular, we use the result from~\cite{9264161} where it is shown that the cosine similarity metric between the user-$i$ of the $k$-th user pair with channel $\mathbf{h}_{k,i}$ and a beam-$b$ with precoding vector $\mathbf{w}_{b}$ can be expressed as follows:
\begin{equation} \label{eq:cosS}
\begin{split}
\cos(\mathbf{h}_{k,i},\mathbf{w}_{b}) & = \frac{\mid \mathbf{h}_{k,i}^{H} \mathbf{w}_{b} \mid}{||\mathbf{h}_{k,i}||~||\mathbf{w}_{b}||}=
      \frac{\mid \boldsymbol{a}(\phi_{k,i})^{H} \boldsymbol{a}(\phi_{b}) \mid}{N} \\ &= F_N \big(\pi [\phi_{k,i} - \phi_b]\big),
\end{split}
\end{equation}
\noindent where $\phi_{k,i} = \sin(\theta_{k,i})$ and $\phi_b=\sin(\bar\theta_b)$ are the normalized directions of the user channel and the candidate beam, respectively, and $F_N$ represents the Fejer Kernel. The properties of Fejer Kernel dictate that as $|\phi_{k,i} - \phi_b|$ increases, $\cos(\mathbf{h}_{{k,i}},\mathbf{w}_{b})\rightarrow 0$. In other words, if a beam and a cooperating cell-center user direction are well aligned, the cosine similarity metric is high and it reflects that it is suitable to schedule this user on that beam. This process is done to schedule each cooperating cell-center user with its best \ac{THz} beam.


\subsection{Cooperating Cell-center and Cell-edge User Pairing: Hungarian Algorithm}
\label{subsection: Cell-cener and Cell-edge User Pairing: Hungarian Algorithm}

At this stage, the \ac{BS} knows the best \ac{THz} beam for scheduling each cooperating cell-center user. It is now the time to pair each cooperating cell-center user with the cell-edge user that has the shortest Euclidean distance to it. To achieve this, a well-known one-to-one assignment matching algorithm entitled the Hungarian method~\cite{kuhn1955hungarian} is adopted. Noting that the computational complexity of the Hungarian method here is $O(K_\textnormal{Coop}^3)$. This computational complexity is significantly simpler than the computational complexity of the brute-force algorithm, i.e., $O(K_\textnormal{Coop}\,!)$.

\begin{algorithm}[!b] 
\caption{A summary of the proposed cooperative scheme} \label{Tab: A summary of the proposed cooperative scheme}
\justify
\noindent~(-)~\textbf{Initialization:} (i) distribute the cooperating cell-center and cell-edge users, (ii) calculate the channel gain of each cooperating cell-center user, (iii) build the \ac{BS} available beams set\;
\noindent~(A)~\textbf{Cooperating cell-center user scheduling stage:} find the best beam for each cooperating cell-center user based on analog beamforming with the aid of the cosine similarity metric\;
\noindent~(B)~\textbf{Cooperating cell-center and cell-edge user pairing stage:} (i) pair each cooperating cell-center user with each cell-edge user based on the Euclidean distance, (ii) in each formed pair, calculate the channel gain between the paired cooperating cell-center user and the cell-edge user\;
\noindent~(C)~\textbf{Power optimization stage:} In each pair, sequentially, (i) calculate the cooperation power of the cooperating cell-center user, and (ii) calculate \ac{BS} \ac{NOMA} power fractions for the cooperating cell-center user and the cell-edge user\;
\end{algorithm}

\subsection{Power Optimization Stage: Sequential Optimization of NOMA BS Power Coefficient and cooperating Cell-center Users Cooperation Power}
\label{subsection: Sequential Optimization of NOMA BS Power Coefficient and Cell-center Users Cooperation Power}

At this stage, the \ac{BS} has calculated the user pairs for all users in the network. Now, we calculate both the cooperation power of each \ac{THz} cooperating cell-center user and the \ac{BS} \ac{NOMA} power fractions. In each user pair, the cooperation power of the \ac{THz} cooperating cell-center user has to be determined such that the minimum required rate, $R_{0}^{\textnormal{min}}$, of the cell-edge user in a user pair is met. This can be achieved by equating $\gamma_{k,j}^{j,\textnormal{Edge}}$ with $\gamma_{0}^{\textnormal{min}}$. Noting that $\gamma_{0}^{\textnormal{min}} = 2^{(R_{0}^{\textnormal{min}}/W)} - 1$. By doing this, $p_{k,u}$ for the $k$-th user pair can be obtained as   
\begin{equation} \label{eq: Cooperation power}
p_{k,u} = \frac{\gamma_{k,j}^{j,\textnormal{Edge}} \sigma_{k,j}^2}{|h_{k,ij}|^2}, \ \forall k \in \textnormal{all user pairs}.
\end{equation}

Next, the \ac{BS} \ac{NOMA} power fractions for the cooperating cell-center user and the cell-edge user in a user pair have to be determined such that we guarantee a successful \ac{SIC} decoding of cell-edge user signal at the cooperating cell-center user. Consequently, in each user pair, we equate $\gamma_{k,j}^{i,\textnormal{Coop}}$ with $\gamma_{k,j}^{j,\textnormal{Edge}}$. After some simple manipulations, $\beta_{k,i}$ (eventually $\beta_{k,j}$ that is equal to $1 - \beta_{k,i}$) can be obtained as \begin{multline} \label{eq: power fractions}
  \beta_{k,i} = \\ \frac{p_{k,u}^2 \kappa |h_{k,ii}|^2 |h_{k,ij}|^2 + \sigma_{k,i}^2 p_{k,u} |h_{k,ij}|^2 - p_k |\mathbf{h}_{k,i}^{H} \mathbf{w}_{k} |^2  \sigma_{k,j}^2}{ - p_k |\mathbf{h}_{k,i}^{H} \mathbf{w}_{k} |^2 \sigma_{k,j}^2 - p_k p_{k,u} |h_{k,ij}|^2 |\mathbf{h}_{k,i}^{H} \mathbf{w}_{k} |^2 },\\
  \forall k \in \textnormal{all user pairs}.
\end{multline}

It is worthy to highlight here how each stage in the proposed scheme reduces the consumed power in the network, which makes this scheme energy-efficient: (A) in stage one, when the cooperating cell-center user in each pair is scheduled in its best beam, this reduces the \ac{BS} transmit power consumption, (B) in stage two, pairing the users based on the Euclidean distance will reduce the required cooperation power at the cooperating cell-center users, and (C) in stage three, optimizing the cooperation power just to satisfy the minimum required rate for the cooperating cell-edge users will reduce the cooperation power consumption. 

\begin{table}[!b]
\centering
\vspace{-1.5em}
\caption{Simulation Parameters}
\label{Tab: Simulation Parameters}
\begin{tabular}{|m{5.0cm}|m{2.9cm}|}
\hline
Parameter name, notation & Value \\ \hline

\ac{BS} transmit power per-channel use, $p_k$ & $[1,3,5,7,9]$ Watt~\cite{9115278}\\
Power consumption of base-band, $P_B$ & 200 mWatt~\cite{9115278}\\
Power consumption of \ac{RF} chain, $P_R$ & 160 mWatt~\cite{9115278}\\
Power consumption of each phase shifter, $P_P$& 40 mWatt~\cite{9115278} \\ 
Power consumption of each power amplifier, $P_A$ & 20 mWatt~\cite{9115278}\\
Inefficiency of power amplifier, $\xi$ & $1/0.38$~\cite{9115278}\\
Users minimum required rate, $R_{0}^{\textnormal{min}}$ & $[5,10,15,20]$ Gbps \cite{8824971}\\ 
Number of antennas at BS, $N$ & $4$~\cite{9264161}\\ 
\ac{BS} antenna gains, $\Omega_{\textnormal{BS}}$ & 20 dBi~\cite{7436794}\\
Cell-center user antenna gain, $\Omega_{\textnormal{User}}$ & $3$ dBi~\cite{7447666}\\
Number of available beams, $B$ & $20$ \\
Beam-width angles, $\bar\theta_b$ & $6^{\circ}$~\cite{9295330}\\
\ac{BS} sector angular coverage, $\bar\theta$ &$120^{\circ}$ \\
Self-interference parameter, $\kappa$ &$0.4$\\
Center frequency of the considered \ac{THz} window, $f$ & $3.42$ THz~\cite{singh2020analytical,9115278} \\
Contiguous Bandwidth of the considered \ac{THz} window, $W$ & $137$ GHz~\cite{singh2020analytical}\\
water vapor molecules absorption coefficient, $k_{\textnormal{abs}}(f)$ & $0.28 \  \textnormal{m}^{-1}$~\cite{gordon2017hitran2016}\\
Number of users, $K_\textnormal{Coop}+K_\textnormal{Edge}$ & $4-20$ users~\cite{9115278}\\
User distribution & Uniform random~\cite{9264161}\\
\ac{BS} coverage region, $d_{\textnormal{Center}}+d_{\textnormal{Edge}} $ & $7$ meters \\
\hline
\end{tabular}
\end{table}


\begin{figure}[!t]
    \centering
    \vspace*{-.02in}
    \subfloat[]{
        \hspace*{-.15in}
        \includegraphics[scale=0.35]{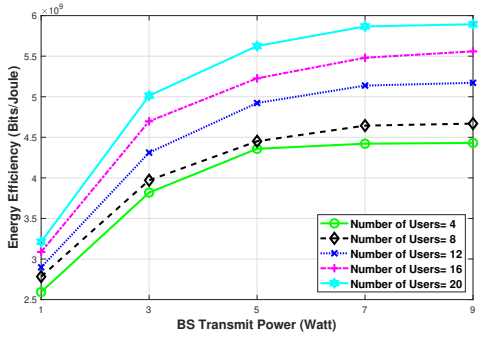}
        \label{fig:EE}
    }
    \hfill
    \subfloat[]{
        \hspace*{-.15in}
        \includegraphics[scale=0.35]{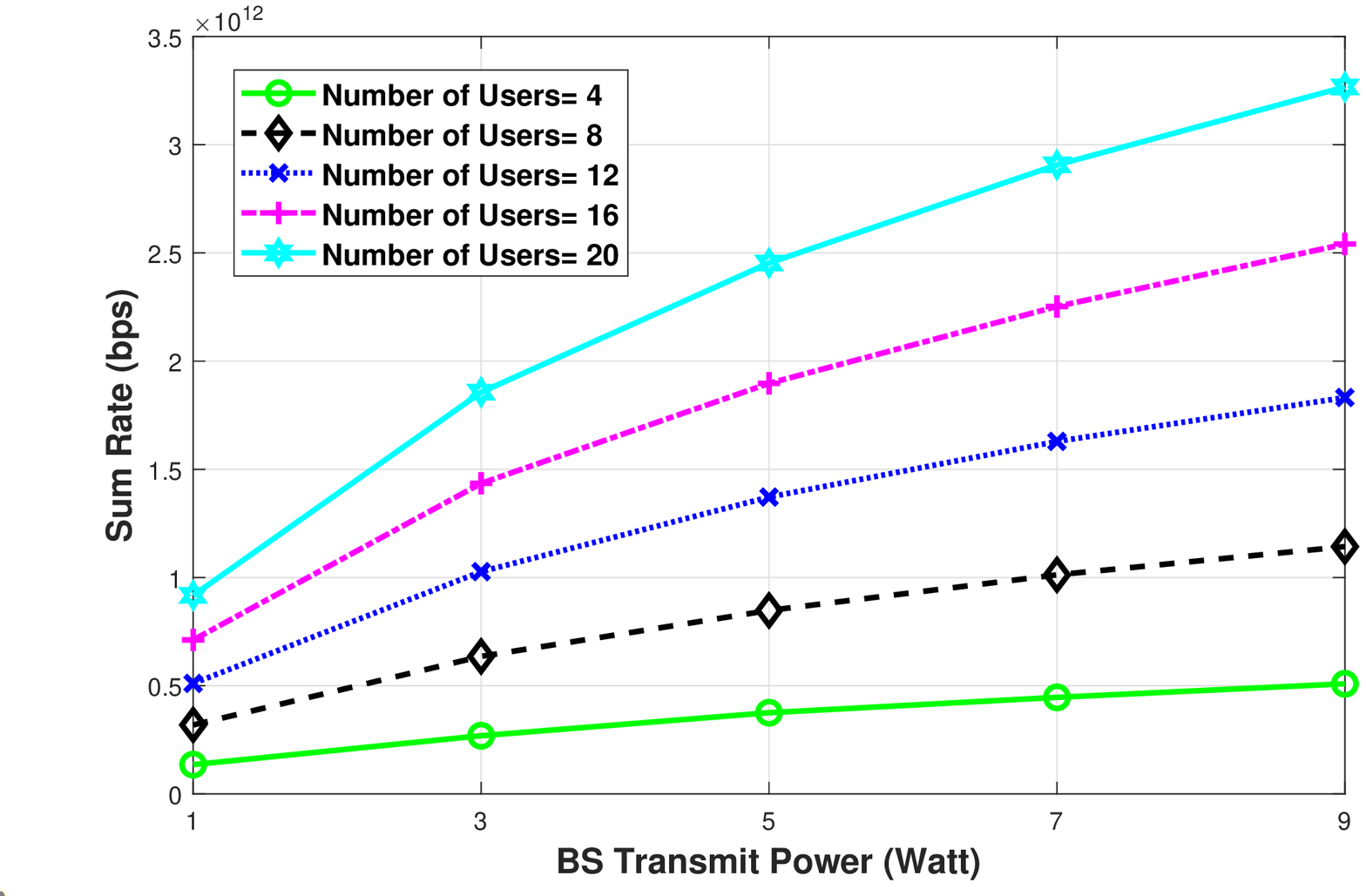}
        \label{fig:SR}
    }
    \hfill
    \subfloat[]{
        \hspace*{-.15in}
        \includegraphics[scale=0.35]{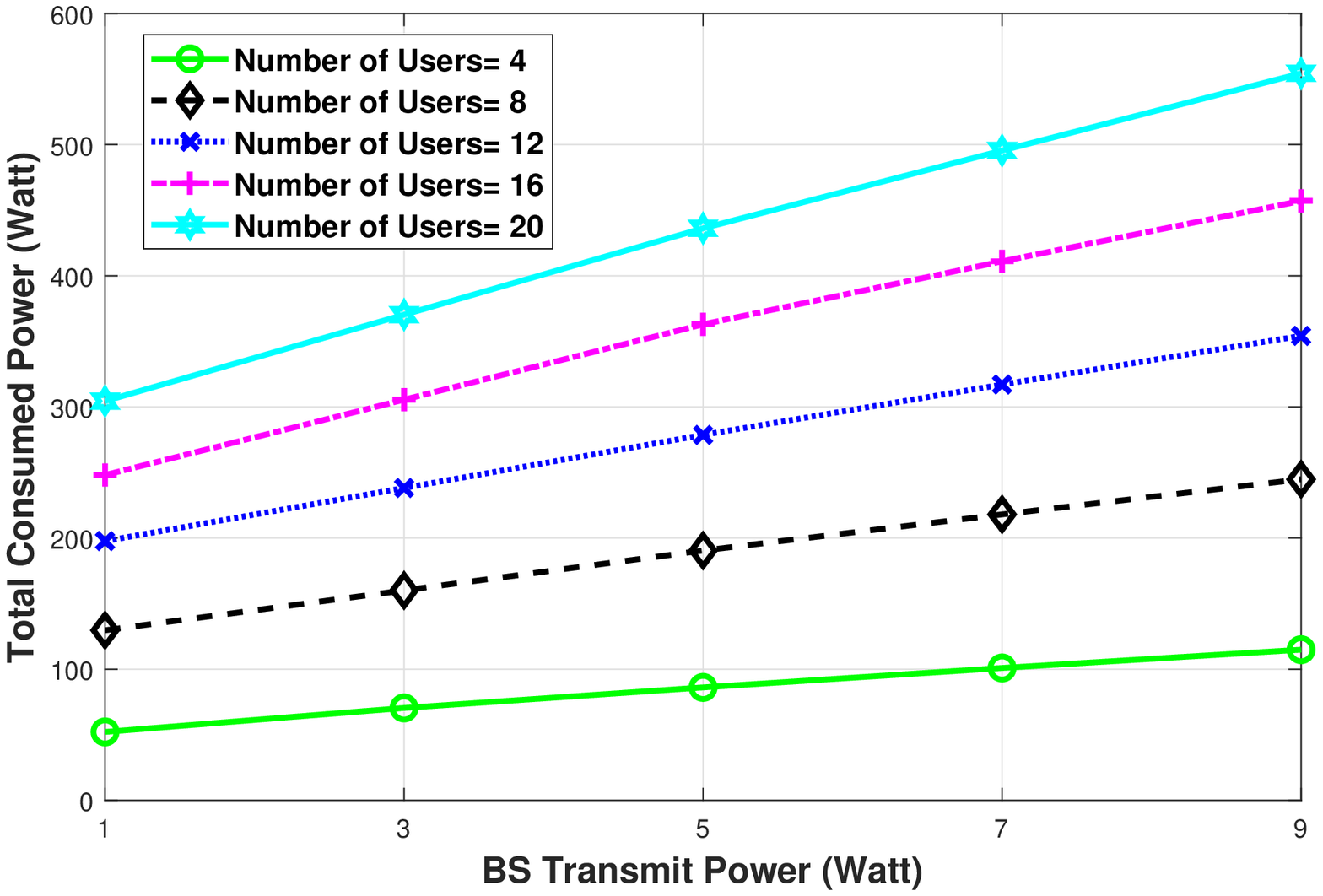}
        \label{fig:ConsumedPower}
    }
    \caption{(a) The energy efficiency, (b) sum-rate, and (c) network-consumed-power performance of the network with different \ac{BS} transmit power levels.}
\label{fig: power levels}
\end{figure}

\begin{figure}[!t]
    \centering
    \vspace*{-.02in}
    \subfloat[]{
        \hspace*{-.1in}
        \includegraphics[scale=0.35]{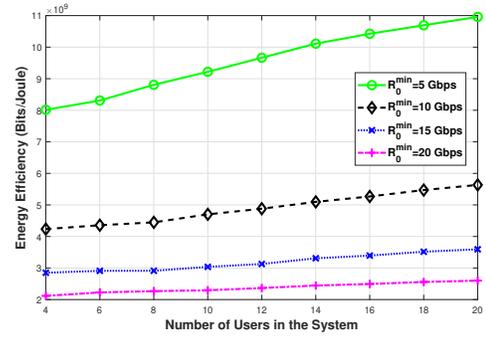}
        \label{fig: changing R0 EE}
    }
    \hfill
  \subfloat[]{
        \hspace*{-.1in}
        \includegraphics[scale=0.35]{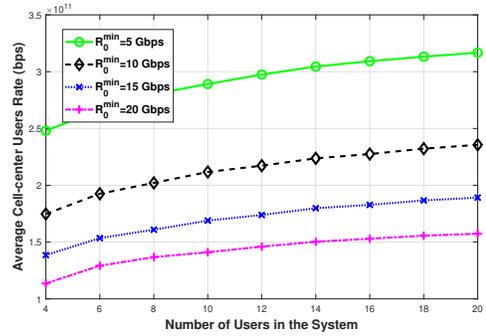}
        \label{fig: changing Cell-center}
    }
    \caption{(a) The achievable \ac{EE}, and (b) the average cooperating cell-center users' rate performance of the network while changing the cell-edge user minimum required rate, $R_{0}^{\textnormal{min}}$.}
\label{fig: minimum user required rate}
\end{figure}

\section{Simulation Results and Discussions}
\label{Section:Results and discussions}

In this section, we evaluate the performance of the proposed cooperative scheme through MATLAB simulations, using the system parameters in Table~\ref{Tab: Simulation Parameters}. Monte-Carlo simulations that are averaged over $10^4$ users’ locations realizations are used for each point in the performance curves. Beyond the listed parameters in Table~\ref{Tab: Simulation Parameters}, the noise power is $\sigma^2=10\,\textnormal{log}_{10}(W)+~N_f-~174$ dBm, where the noise figure $N_f=10$ dB~\cite{9264161}. Since this work focuses on achieving the minimum required rate for the cell-edge users, we assume that the subset of the cell-center users that are capable of helping the cell-edge users, with the minimum required cooperation power and hence the better energy efficiency of the proposed scheme, are the ones that are located in the farthest $20\%$ of cell-center users coverage region. Such scenario can take place in exhibition halls, conferences, restaurants, etc. that most likely have dense user deployment. 


Fig.~\ref{fig: power levels} illustrates the energy efficiency, sum-rate, and network-consumed-power performance of the proposed cooperative scheme with different \ac{BS} transmit power levels and with different number of users in the network. The first observation here, is that as the \ac{BS} transmit power increases the network \ac{EE} increases then starts to saturate as can be seen in Fig.~\ref{fig:EE}. Such a trend in the network 
can be justified by the following argument; the increase in the \ac{BS} transmit power is reflected as (i) logarithmic increase in the network sum rate (Fig.~\ref{fig:SR}) and as (ii) linear increase in the total network consumed power (Fig.~\ref{fig:ConsumedPower}). The second observation here, is that as the number of considered users in the network increases both the network \ac{EE} and the users' sum rate increase; this is evident in the \ac{EE} formula~\eqref{eq: GEE formula}.

\begin{figure}[!t]
    \centering
    \vspace*{-.02in}
    \subfloat[mmWave-NOMA counterpart scheme.]{
        \hspace*{-.1in}
        \includegraphics[scale=0.35]{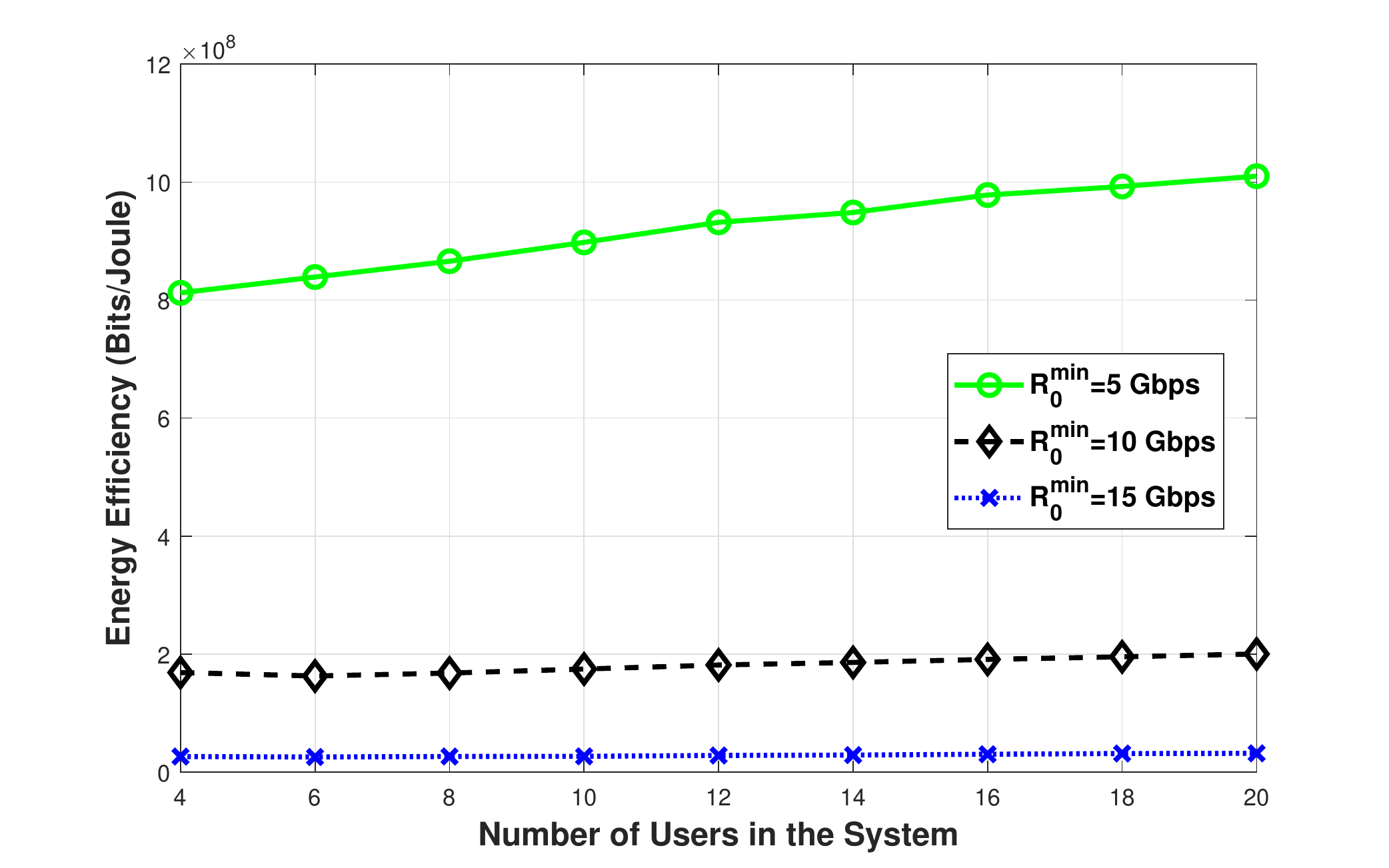}
        \label{fig: changing R0 EE mmWave}
    }
    \hfill
  \subfloat[mmWave-NOMA counterpart scheme.]{
        \hspace*{-.1in}
        \includegraphics[scale=0.35]{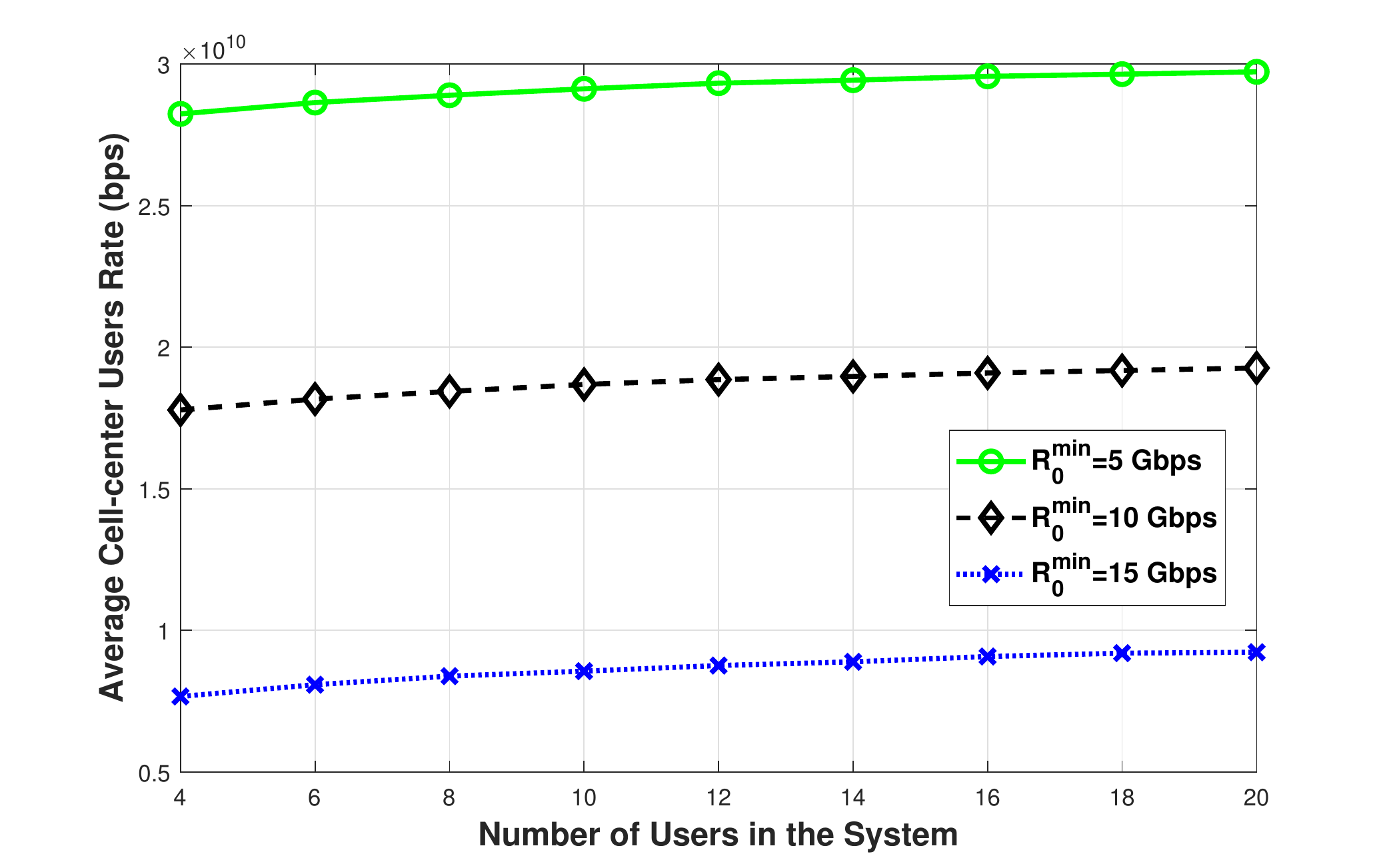}
        \label{fig: changing Cell-center mmWave}
    }
    \caption{(a) The achievable \ac{EE}, and (b) the average cooperating cell-center users' rate performance of mmWave-NOMA counterpart scheme while changing cell-edge user minimum required rate, $R_{0}^{\textnormal{min}}$.}
\label{fig: minimum user required rate mmWave}
\end{figure}

In Fig.~\ref{fig: changing R0 EE}, we present the achievable \ac{EE} performance of the system while changing the minimum user required rate, $R_{0}^{\textnormal{min}}=[5,10,15,20]$ Gbps. We provide this figure to support the effectiveness of the proposed cooperative scheme in improving the network \ac{EE} while guaranteeing a minimum rate requirement for the cell-edge users. From this figure, one can see that as the minimum user required rate increases the network \ac{EE} performance decreases; this is because with this increase in the minimum user required rate, the cooperation power at the cooperating cell-center users increases, consequently, the total consumed power in the system increases and the network \ac{EE} of the system decreases. In Fig.~\ref{fig: changing Cell-center}, we provide the average cooperating cell-center users' rate performance of the network while changing $R_{0}^{\textnormal{min}}$. This figure demonstrates the ability of the proposed cooperative scheme in achieving high data rate for the cooperating cell-center users, which are between $0.1$-$0.3$ Tbps, while guaranteeing a minimum user required rate for all the cell-edge users. For benchmarking purposes, in Fig.~\ref{fig: minimum user required rate mmWave}, we provide the achievable \ac{EE} and the average cooperating cell-center users' rate performance of the mmWave-NOMA counterpart scheme. We assume that the mmWave carrier frequency is $f_{\textnormal{mmWave}}=28$ GHz with a contiguous bandwidth of $W_{\textnormal{mmWave}}=2$ GHz~\cite{8454272}. Comparing Fig.~\ref{fig: minimum user required rate} with Fig.~\ref{fig: minimum user required rate mmWave}, it is evident that both the achievable \ac{EE} and average cooperating cell-center users' rate have a large gain as a consequence of operating in the THz band compared with the mmWave band.

\section{Conclusions and Future Research Directions}
\label{Section:Conclusions}

As both operating at the \ac{THz} bands and \ac{NOMA}-enabled scheme are envisioned to be among the enablers of high data-rate applications in future wireless networks, an energy-efficient scheme of multi-user indoor \ac{THz}-\ac{MISO} \ac{NOMA}-enabled system with user-assisted cooperative side links is proposed and its performance investigated in this paper. Analog beamforming with the aid of the cosine similarity metric has been adopted to allocate cell-center users to \ac{BS} \ac{THz} beams, then the user-pairing problem between cooperating cell-center users and cell-edge users has been tackled through the Hungarian algorithm that pairs users based on the Euclidean distance, finally, the power allocation of the \ac{BS} transmit power as well as the cooperation power of the cooperating cell-center users has been sequentially optimized. The obtained results demonstrate the energy efficiency and illustrate the effect of changing different network parameters, such as, changing the \ac{BS} transmit power levels and changing cell-edge users' minimum required rate on the \ac{EE} of the network. We intend to extend this work by taking into account some practical considerations such as the performance of the proposed scheme with imperfect \ac{CSI} and with \ac{NLoS} scenarios. 

\section{Acknowledgement}
\label{Section:Acknowledgement}
This work was supported in part by the King Fahd University of Petroleum and Minerals under Grant SB191038; in part by Ericsson Canada Inc.; and in part by the Discovery Grant of the Natural Sciences and Engineering Research Council of Canada.

\bibliographystyle{IEEEtran}
\bibliography{main}


\end{document}

%% file: glossaries.tex
\newacronym{SA}{SA}{Simulated Annealing}
\newacronym{VLC}{VLC}{visible light communications}
\newacronym{RF}{RF}{radio frequency}
\newacronym{V2V}{V2V}{vehicle-to-vehicle}
\newacronym{V2X}{V2X}{vehicle-to-everything}
\newacronym{B5G}{B5G}{beyond-fifth generation}
\newacronym{5G}{5G}{fifth-generation}
\newacronym{6G}{6G}{sixth-generation}
\newacronym{LED}{LED}{light emitting diode}
\newacronym{LEDs}{LEDs}{light emitting diodes}
\newacronym{OMA}{OMA}{orthogonal multiple access}
\newacronym{FDMA}{FDMA}{frequency-division multiple-access}
\newacronym{TDMA}{TDMA}{time-division multiple-access}
\newacronym{CDMA}{CDMA}{code-division multiple-access}
\newacronym{OFDMA}{OFDMA}{orthogonal frequency-division multiple-access}
\newacronym{WDMA}{WDMA}{wavelength-division multiple-access}
\newacronym{NOMA}{NOMA}{Non-orthogonal multiple access}
\newacronym{PD-NOMA}{PD-NOMA}{power-domain NOMA}
\newacronym{SC}{SC}{superposition coding}
\newacronym{SIC}{SIC}{successive interference cancellation}
\newacronym{BS}{BS}{base station}
\newacronym{QoS}{QoS}{quality-of-service}
\newacronym{NP}{NP}{non-deterministic polynomial-time}
\newacronym{DCO-OFDM}{DCO-OFDM}{direct-current biased optical-OFDM}
\newacronym{ITU}{ITU}{international telecommunication union}
\newacronym{FoV}{FoV}{field-of-view}
\newacronym{FoVs}{FoVs}{field-of-views}
\newacronym{CSI}{CSI}{channel state information}
\newacronym{LACO-OFDM}{LACO-OFDM}{layered asymmetrically clipped optical OFDM}
\newacronym{FR}{FR}{frequency reuse}
\newacronym{EAs}{EAs}{evolutionary algorithms}
\newacronym{C-LiAN}{C-LiAN}{centralized light access network}
\newacronym{AP}{AP}{access point}
\newacronym{APs}{APs}{access points}
\newacronym{PD}{PD}{photo-diode}
 \newacronym{PDs}{PDs}{photo-diodes}
\newacronym{SINR}{SINR}{signal-to-noise-interference ratio}
\newacronym{LoS}{LoS}{line-of-sight}
\newacronym{AWGN}{AWGN}{additive white Gaussian noise}
\newacronym{SNR}{SNR}{signal-to-noise ratio}
\newacronym{NLUPA}{NLUPA}{next-largest-difference user-pairing algorithm}
\newacronym{D-NLUPA}{D-NLUPA}{divide-and-next-largest-difference user-pairing algorithm}
\newacronym{NAICS}{NAICS}{network-assisted interference cancellation and suppression}
\newacronym{LTE}{LTE}{long term evolution}
\newacronym{3GPP}{3GPP}{3rd Generation Partnership Project}
\newacronym{CR}{CR}{cognitive radio}
\newacronym{2D}{2D}{two-dimension}
\newacronym{GP}{GP}{gradient projection}
\newacronym{umMTC}{umMTC}{ultra-massive machine-type communication}
\newacronym{IoE}{IoE}{internet-of-everything}
\newacronym{ZF}{ZF}{zero-forcing}
\newacronym{NLIP}{NLIP}{non-linear integer programming}
\newacronym{DP}{DP}{dynamic programming}
\newacronym{THz}{THz}{Terahertz}
\newacronym{D2D}{D2D}{device-to-device}
\newacronym{mmWave}{mmWave}{Millimeter-wave}
\newacronym{PS}{PS}{phase shifter}
\newacronym{PA}{PA}{power amplifier}
\newacronym{ABF}{ABF}{analog beamforming}
\newacronym{SPS}{SPS}{single-phase shifter}
\newacronym{AWV}{AWV}{antenna weight vector}
\newacronym{EE}{EE}{energy efficiency}
\newacronym{SE}{SE}{spectral efficiency}
\newacronym{NLoS}{NLoS}{non-line-of-sight}
\newacronym{mMIMO}{mMIMO}{massive multiple-input multiple-output}
\newacronym{MISO}{MISO}{multiple-input-single-output}
\newacronym{SISO}{SISO}{single-input single-output}
\newacronym{MOO}{MOO}{multi-objective optimization}
\newacronym{HP}{HP}{hybrid beamforming}
\newacronym{SCA}{SCA}{successive convex approximation }
\newacronym{SOC}{SOC}{second order cone}
\newacronym{RHS}{RHS}{right-hand side}
\newacronym{SDR}{SDR}{semidefinite relaxation}
\newacronym{VR}{VR}{virtual reality}
\newacronym{XR}{XR}{augmented reality}
\newacronym{NR}{NR}{new radio}
\newacronym{FD}{FD}{full-duplex}
\newacronym{QoE}{QoE}{quality-of-experience}
\newacronym{ULA}{ULA}{uniform linear array}
\newacronym{SI}{SI}{self-interference}
\newacronym{DF}{DF}{decode-and-forward}
\newacronym{GEE}{GEE}{global energy efficiency}
\newacronym{CNOMA}{CNOMA}{cooperative NOMA}